\documentclass[sigconf]{acmart}
\AtBeginDocument{%
  }

\usepackage{subcaption}   
\usepackage{graphicx} 
\usepackage{amsmath}
\usepackage{tabularx}
\usepackage{multirow}

\begin{document}

\title{RecBundle: A Next-Generation Geometric Paradigm for Explainable Recommender Systems}


\author{Hui Wang}
\authornote{Both authors contributed equally to this research.}
\email{wanghui4042@iie.ac.cn}
\affiliation{%
  \institution{Institute of Information Engineering, Chinese Academy of Sciences}
  \city{Beijing}
  \country{China}
}

\author{Tianzhu Hu}
\authornotemark[1]
\email{hutianzhu@iie.ac.cn}
\affiliation{%
  \institution{Institute of Information Engineering, Chinese Academy of Sciences}
  \city{Beijing}
  \country{China}
}

\author{Mingming Li}
\authornote{Corresponding authors.}
\email{limingming@iie.ac.cn}
\affiliation{%
  \institution{Institute of Information Engineering, Chinese Academy of Sciences}
  \city{Beijing}
  \country{China}
}

\author{Xi Zhou}
\email{zhouxi@iie.ac.cn}
\affiliation{%
  \institution{Institute of Information Engineering, Chinese Academy of Sciences}
  \city{Beijing}
  \country{China}
}

\author{Chun Gan}
\authornotemark[2]
\email{ganchun@jd.com}
\affiliation{%
  \institution{JD.com}
  \city{Beijing}
  \country{China}
}

\author{Jiao Dai}
\email{daijiao@iie.ac.cn}
\affiliation{%
  \institution{Institute of Information Engineering, Chinese Academy of Sciences}
  \city{Beijing}
  \country{China}
}

\author{Jizhong Han}
\email{hanjizhong@iie.ac.cn}
\affiliation{%
  \institution{Institute of Information Engineering, Chinese Academy of Sciences}
  \city{Beijing}
  \country{China}
}

\author{Songlin Hu}
\email{husonglin@iie.ac.cn}
\affiliation{%
  \institution{Institute of Information Engineering, Chinese Academy of Sciences}
  \city{}
  \country{}
}
\affiliation{%
  \institution{School of Cyberspace Security, University of Chinese Academy of Sciences}
  \city{Beijing}
  \country{China}
}

\author{Tao Guo}
\email{guotao@iie.ac.cn}
\affiliation{%
  \institution{Institute of Information Engineering, Chinese Academy of Sciences}
  \city{Beijing}
  \country{China}
}

% \affiliation{%

% \affiliation{%

%  Institute of Information Engineering, Chinese Academy of Sciences, Beijing, China \\
%  \and School of Cyber Security, University of Chinese Academy of Sciences, Beijing, China


% \renewcommand{\shortauthors}{Wang et al.}

\begin{abstract}
Recommender systems are inherently dynamic feedback loops where prolonged local interactions accumulate into macroscopic structural degradation such as information cocoons. Existing representation learning paradigms are universally constrained by the assumption of a single flat space, forcing topologically grounded user associations and semantically driven historical interactions to be fitted within the same vector space. This excessive coupling of heterogeneous information renders it impossible for researchers to mechanistically distinguish and identify the sources of systemic bias. To overcome this theoretical bottleneck, we introduce Fiber Bundle from modern differential geometry and propose a novel geometric analysis paradigm for recommender systems. This theory naturally decouples the system space into two hierarchical layers: the base manifold formed by user interaction networks, and the fibers attached to individual user nodes that carry their dynamic preferences. Building upon this, we construct RecBundle, a framework oriented toward next-generation recommender systems that formalizes user collaboration as geometric connection and parallel transport on the base manifold, while mapping content evolution to holonomy transformations on fibers. From this foundation, we identify future application directions encompassing quantitative mechanisms for information cocoons and evolutionary bias, geometric meta-theory for adaptive recommendation, and novel inference architectures integrating large language models (LLMs). Empirical analysis on real-world MovieLens and Amazon Beauty datasets validates the effectiveness of this geometric framework.

\end{abstract}






\maketitle

\section{Introduction}

With the deep integration of mobile internet and artificial intelligence, recommender systems have evolved into critical nexuses connecting massive data with individual users. Yet these systems constitute dynamic feedback loops where user behaviors, content attributes, and algorithmic strategies interact and co-evolve over time \cite{piao2023human}. The accumulation of micro-level local decisions through these loops ultimately confronts systems with deep-rooted challenges including interpretability deficits and structural fragility, which macroscopically manifest as information cocoons. Although recent advances in graph neural networks \cite{liu2024rumor}, multi-objective optimization, causal inference, and generative models \cite{peng2024reconciling,wu2024enhancing,zhang2024practical,wang2022user} have achieved remarkable progress, these methods predominantly focus on static preference fitting and observational statistical corrections, lacking a rigorous mathematical characterization of how information propagates and evolves withi the n user-item space.

% With the deep integration of the mobile internet and artificial intelligence, recommender systems have evolved into a critical nexus connecting massive data with individual users. Yet these systems are far from static matching engines; they constitute dynamic feedback loops where user behaviors, content attributes, and algorithmic strategies interact and co-evolve over time \cite{piao2023human}. Over long-term interactions, the accumulation of micro-level local decisions through feedback loops ultimately confronts the system with deep-rooted challenges, such as a lack of interpretability and structural fragility, which macroscopically manifest as the exacerbation of information cocoons. Although recent advances, including high-order topological feature extraction via graph neural networks (GNNs) \cite{liu2024rumor} and the integration of multi-objective optimization, causal inference, and generative models \cite{peng2024reconciling,wu2024enhancing,zhang2024practical,wang2022user} have achieved remarkable progress, these methods predominantly focus on static preference fitting and observational statistical corrections. They largely lack a rigorous mathematical characterization of how information propagates and evolves within the user-item space.

A fundamental limitation underlies existing paradigms. Whether employing Euclidean spaces or recent hyperbolic and Riemannian geometries, they invariably force user preferences and item features into a single global metric space. This inherently couples two heterogeneous mechanisms that drive information flow in recommender systems. The first is population-level collaboration, namely horizontal information diffusion through user connection structures. The second is individual-level evolution, or vertical accumulation of dynamic preferences via semantic continuity of historical interactions. By forcing these heterogeneous signals into identical representation spaces for joint optimization, mainstream approaches make it theoretically impossible to pinpoint the origin of systemic biases when they emerge.

To resolve this, we turn to Fiber Bundle theory \cite{ZRZZ202303006,SciencePress2006}. Widely adopted in gauge theory \cite{liang2023differential} and computer vision \cite{DBLP:conf/iclr/CourtsK22, DBLP:journals/entropy/Nielsen20c}, this theory has proven effective in describing how complex systems assemble local product structures into intricate global topologies \cite{DBLP:journals/ijrr/OrtheyAT24, DBLP:conf/iros/OrtheyEY18, DBLP:conf/isrr/OrtheyT19}. It models the system as a composite object. A base manifold formed by the discrete user interaction graph encodes topological distances and collective collaborative relationships. Fibers attached to individual user nodes carry continuous preference evolution. This orthogonal decoupling separates who resembles whom from how a specific user evolves, enabling separate modeling of collective structure and individual dynamics.

% Decoupling these two heterogeneous signals demands a mathematical framework that accommodates both global structure and local dynamics, which aligns perfectly with the core advantage of Fiber Bundle theory \cite{ZRZZ202303006,SciencePress2006}. Widely adopted in fields such as gauge theory \cite{liang2023differential} and computer vision \cite{DBLP:conf/iclr/CourtsK22, DBLP:journals/entropy/Nielsen20c}, this theory has proven highly effective in describing how complex systems assemble local product structures into intricate global topologies \cite{DBLP:journals/ijrr/OrtheyAT24, DBLP:conf/iros/OrtheyEY18, DBLP:conf/isrr/OrtheyT19}.Under this perspective, recommender systems are no longer reduced to uniform metric spaces but are instead modeled as hierarchical composite geometric objects: the discrete interaction graph spanning all users constitutes the base manifold of the system, encoding topological distances and collective collaborative relationships among users, while the independent semantic spaces attached to individual user nodes, accommodating continuous preference evolution, constitute the corresponding fibers. In essence, the base manifold characterizes who resembles whom, whereas the fibers characterize how a specific user evolves, this natural decoupling enables separate modeling of collective structure and individual dynamics.

Building on this insight, we propose RecBundle, a geometric analysis framework for next-generation explainable recommender systems. Using the language of Fiber Bundle, this framework mathematically reconstructs collaborative and evolutionary mechanisms. It formalizes inter-user information collaboration as geometric connections and parallel transport on the base manifold. It maps long-term preference evolution to holonomy transformations on fiber spaces. This theoretical foundation provides unified mathematical grounding for the following contributions:
% Building upon these theoretical insights, we propose RecBundle, a geometric analysis framework for next-generation recommender systems. By employing the language of fiber bundles, this framework mathematically reconstructs the collaborative and evolutionary mechanisms within the system. It formalizes inter-user information collaboration as geometric connections and parallel transport on the base manifold, and maps the long-term evolution of individual preferences to holonomy transformations on the fiber spaces. This paper focuses on establishing this theoretical framework and, on this basis, provides a unified mathematical foundation for the following three research directions:
\vspace{-0.4em}
\begin{itemize}
\item \textbf{A quantifiable framework for information evolution.} It translates phenomena like information cocoons into geometric anomalies: local curvature distortion and holonomy-induced dimensional contraction, providing rigorous tools for analyzing information degradation.

\item \textbf{A geometric meta-theory for adaptive recommendation.} It reconceptualizes personalization as geometric construction, where algorithms dynamically match manifold structures to each user's cognitive state.

\item \textbf{A novel reasoning architecture integrating Large Language Models.} It maps fast-and-slow thinking onto Fiber Bundle geometry, using cross-node retrieval for rapid alignment and fiber constraints to ensure semantic consistency in chain-of-thought reasoning.
\end{itemize}

\section{Preliminaries}
This section briefly reviews the Fiber Bundle theory, providing a unified mathematical language for decoupling topological structures and dynamic semantics in recommender systems.

\subsection{Topological Structure}

\begin{figure}[htp]
\centering
\includegraphics[scale=0.45]{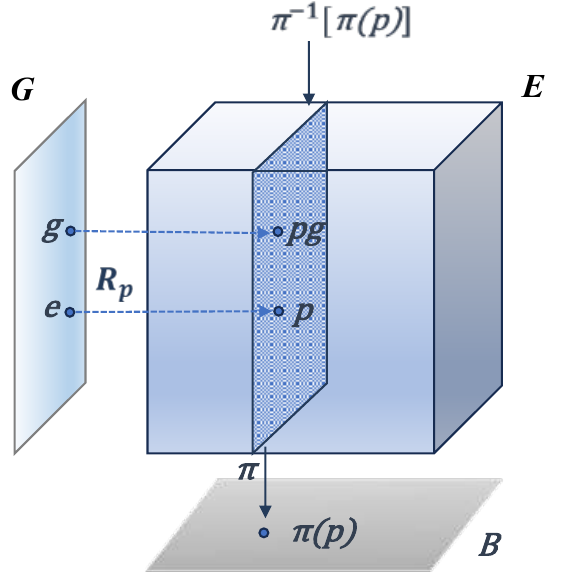}

\caption{Simple Diagram of a Principal Fiber Bundle.} \label{Fig1}
\end{figure}

Mathematically, a fiber bundle \cite{ManifoldHomotopy}
($E$,$B$,$F$,$\pi$) decouples a complex state space into a base manifold $B$ that serves as the topological skeleton whose points represent distinct entities, with attached fibers $F$ that function as independent state spaces characterizing internal dynamics. The projection map $\pi:E \rightarrow B$ anchors high-dimensional states in the total space $E$ to specific entities on $B$, thereby orthogonally decoupling an entity’s physical topology from its internal semantics, as shown in Figure ~\ref{Fig1}.

Specifically, when the fiber itself is a Lie group $G$, this structure constitutes a principal bundle \cite{liang2023differential}, which is rigorously characterized by a free and smooth right group action on the total space, a surjective projection mapping, and equivariant local trivializations that ensure global structural consistency.

% Mathematically, a fiber bundle is defined as a 4-tuple ($E$,$B$,$F$,$\pi$). Its core lies in deconstructing the state space of a complex system into local product spaces. Specifically, the base manifold $B$ constitutes the fundamental topological skeleton of the system, where each point represents an entity with a distinct spatial position. The fiber $F$ is an independent state space attached to each specific point on the base manifold, characterizing the internal dynamic features of that entity. The total space $E$ integrates the base manifold and all fibers. Through the projection map $\pi:P \rightarrow B$, any high-dimensional state in the total space can be strictly anchored to a specific entity on the base manifold, thereby geometrically achieving an orthogonal decoupling between the entity's physical position and its internal semantic state, as illustrated in Figure ~\ref{Fig1}. 

\vspace{-0.8em}

\subsection{Evolution Operators}
Since fibers at different manifold points are mutually independent vector spaces, direct cross-node algebraic operations are geometrically invalid. To enable cross-node information passing, a connection $\omega$ must be introduced, which defines how a fiber state at one point can be parallel transported along a path on the base manifold to the fiber of an adjacent point. However, when the base manifold is non-flat, i.e., possessing non-zero curvature $\Omega$, this parallel transport exhibits strong path dependence. If a state vector is parallel transported along a closed loop $\gamma$ on the base manifold back to its starting point, an irreversible linear transformation typically occurs, known as a holonomy transformation $Hol(\gamma)$. This geometric projection and dimensionality reduction effect induced by closed-loop paths provides a precise mathematical tool for quantifying state contraction in the long-term evolution of complex systems.

\section{RecBundle: A Geometric Analysis Framework}

To introduce the mathematical constructs of Fiber Bundle into the practical modeling of recommender systems, this chapter proposes the RecBundle (Recommendation on Fiber Bundle) analysis framework. This framework aims to establish a formal correspondence between the information passing mechanisms of recommender systems and manifold geometric operators. From this perspective, we provide theoretical support for quantifying macro-level biases during the system's dynamic evolution.

\subsection{Geometric Mapping of the Dual Collaborative Mechanisms}

\begin{figure}[t]  % 跨栏浮动体
\centering
\begin{subfigure}[b]{0.48\columnwidth}
  \centering
  \includegraphics[width=1\textwidth]{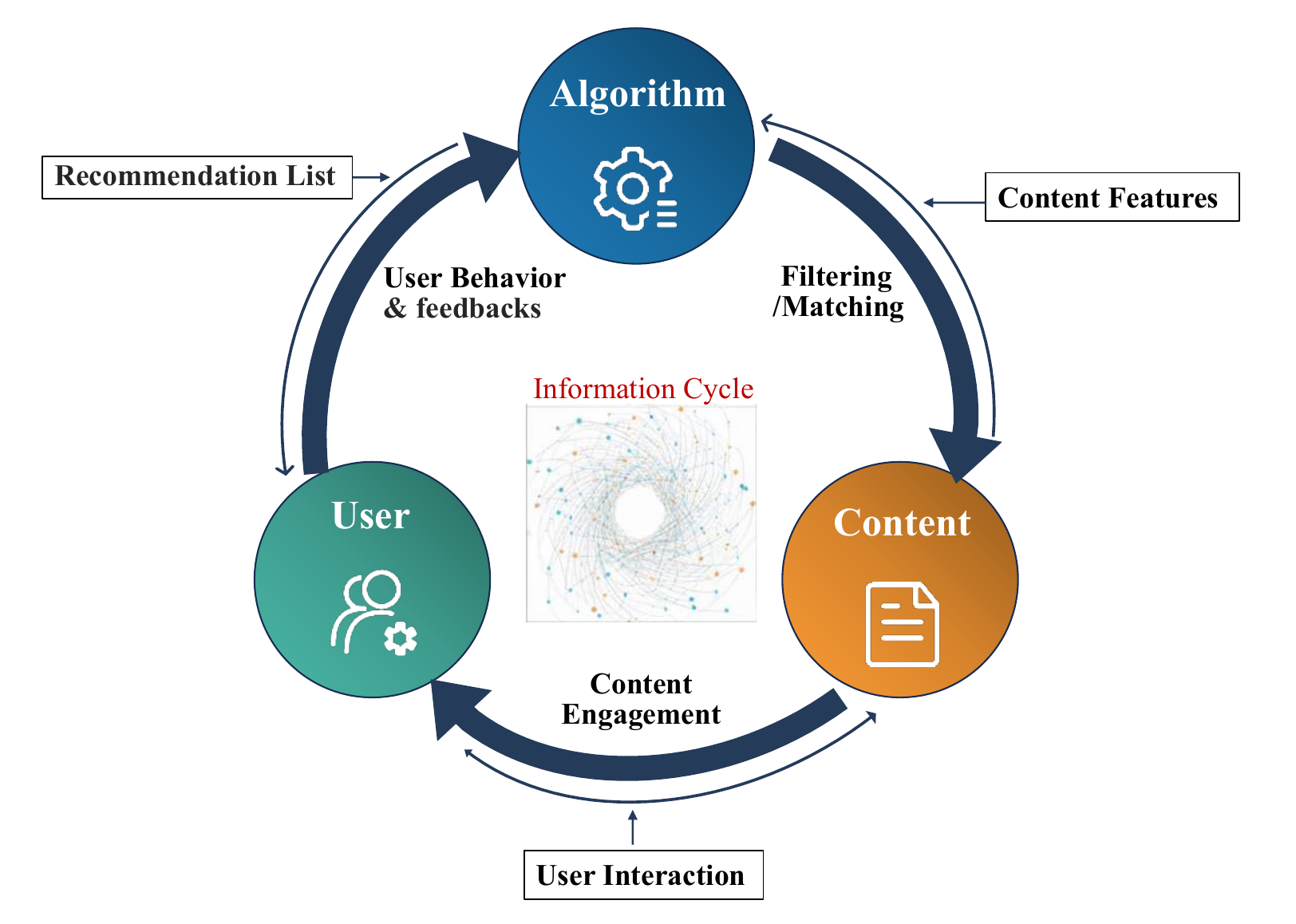}
  \caption{Information Flow(Macro)}
  \label{fig:3a}
\end{subfigure}\hfill
\begin{subfigure}[b]{0.48\columnwidth}
  \centering
  \includegraphics[width=0.7\textwidth]{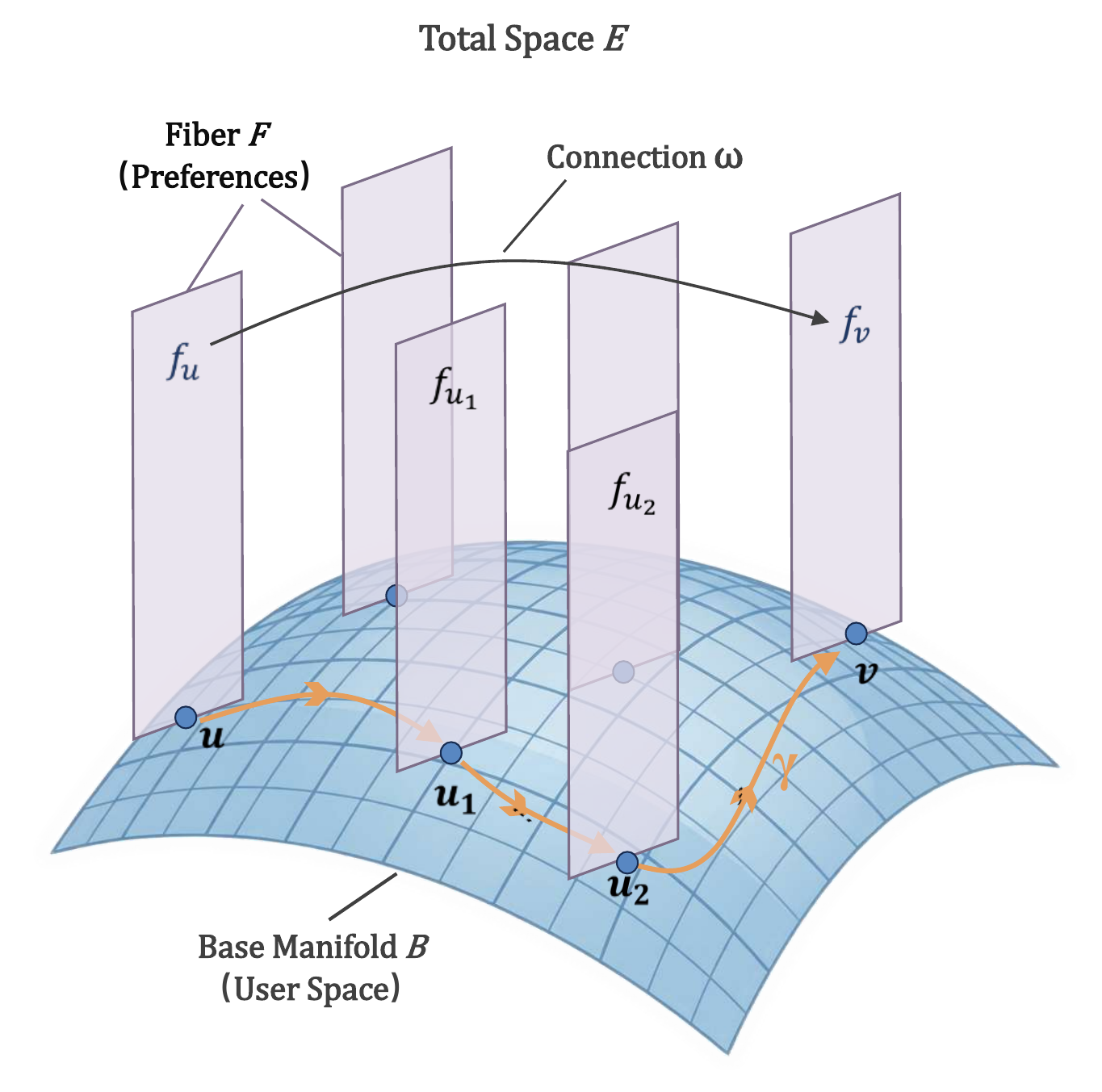}
  \caption{Fiber Bundle Formalization}
  \label{fig:3b}
\end{subfigure}
\caption{Paradigm Shift in Recommender Systems Modeling. }
% (a) The traditional macroscopic feedback loop among users, algorithms, and content. (b) The RecBundle framework, which orthogonally decouples user topology (base manifold) from dynamic semantic states (fibers) to isolate heterogeneous evolutionary mechanisms.
\label{fig:3}
\end{figure}

In the RecBundle framework, we bridge the macroscopic information cycle of recommender systems with rigorous mathematical constructs, as illustrated in Figure~\ref{fig:3a}. Specifically, we formalize the correspondence between these core system elements and the geometric objects of Fiber Bundle, as shown in Figure~\ref{fig:3b}. As summarized in Table~\ref{tab:1} and ~\ref{tab:2}, this geometric mapping mathematically decouples the recommendation process. The base manifold captures the collaborative user topology, while the attached fibers model dynamic preference updates. This formulation not only unifies mainstream recommendation paradigms as special geometric cases, but also introduces curvature and holonomy as concrete metrics to quantify system biases and structural vulnerabilities.

% As summarized in Table~\ref{tab:1}, this mapping naturally decouples the previously mentioned heterogeneous information: the base manifold governs horizontal population collaboration, the fibers correspond to individual vertical content evolution, while geometric curvature and holonomy transformations serve as computable metrics for analyzing system biases.

\begin{table*}
  \caption{Correspondence between Recommender System Concepts and Fiber Bundle Theory}
  \small
  \label{tab:1}
  \begin{tabular}{@{}p{3.0cm}p{3.5cm}>{\centering\arraybackslash}m{1.4cm}p{7.3cm}@{}}
    \toprule
   RS Concept & Fiber Bundle  Object & Symbol & Practical Meaning \& Function\\
    \midrule
    User Set \& Topology & Base Space & $B(U)$ & Network topology encoding user cognitive distances.\\
    Preferences  Space & Fiber & $F_u$  & Independent semantic spaces for dynamic user interests.\\
    Sequential Interactions & Path on Manifold & $\gamma(t)$ & Historical interaction trajectories or feedback loops.\\
    Feature Alignment & Connection & $\omega$  & Geometric rules for cross-node feature alignment.\\
    Collaborative & Parallel Transport & $P_{u \rightarrow v}$  & Projection operators for neighbor feature aggregation.\\
    Local Bias & Connection Curvature & $\Omega$ & Spatial distortions during aggregation.\\
   Long-term Evolution & Holonomy Transformation & $Hol(\gamma)$  & Closed-loop evolving transformations.\\
    \bottomrule
  \end{tabular}
\end{table*}

\subsection{Parallel Transport}

\begin{figure*}[t]  % 跨栏浮动体
\centering
\begin{subfigure}[b]{0.5\textwidth}
  \centering
  \includegraphics[width=1\textwidth]{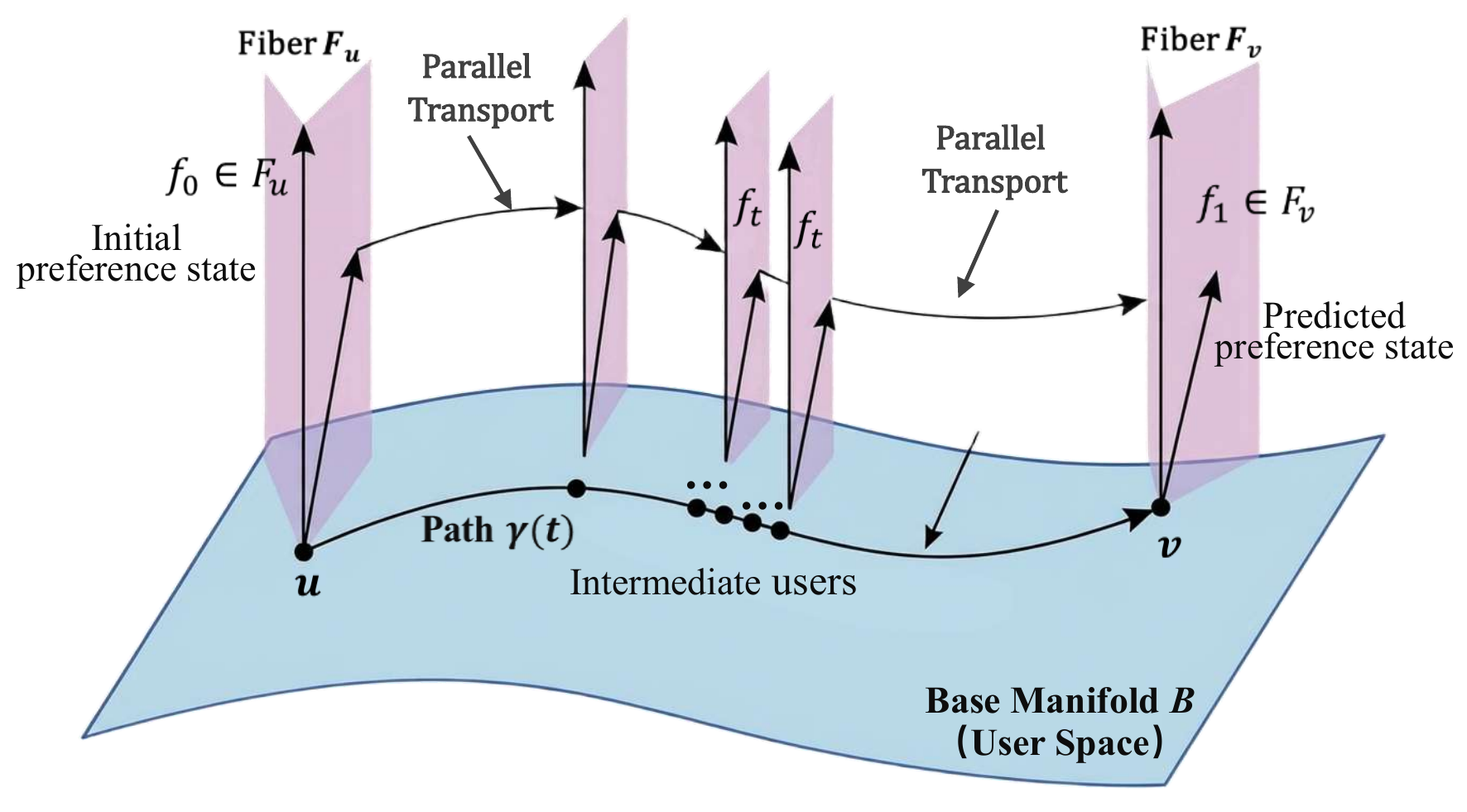}
  \caption{Collaborative aggregation (parallel transport)}
  \label{fig:4a}
\end{subfigure}\hfill
\begin{subfigure}[b]{0.5\textwidth}
  \centering
  \includegraphics[width=0.8\textwidth]{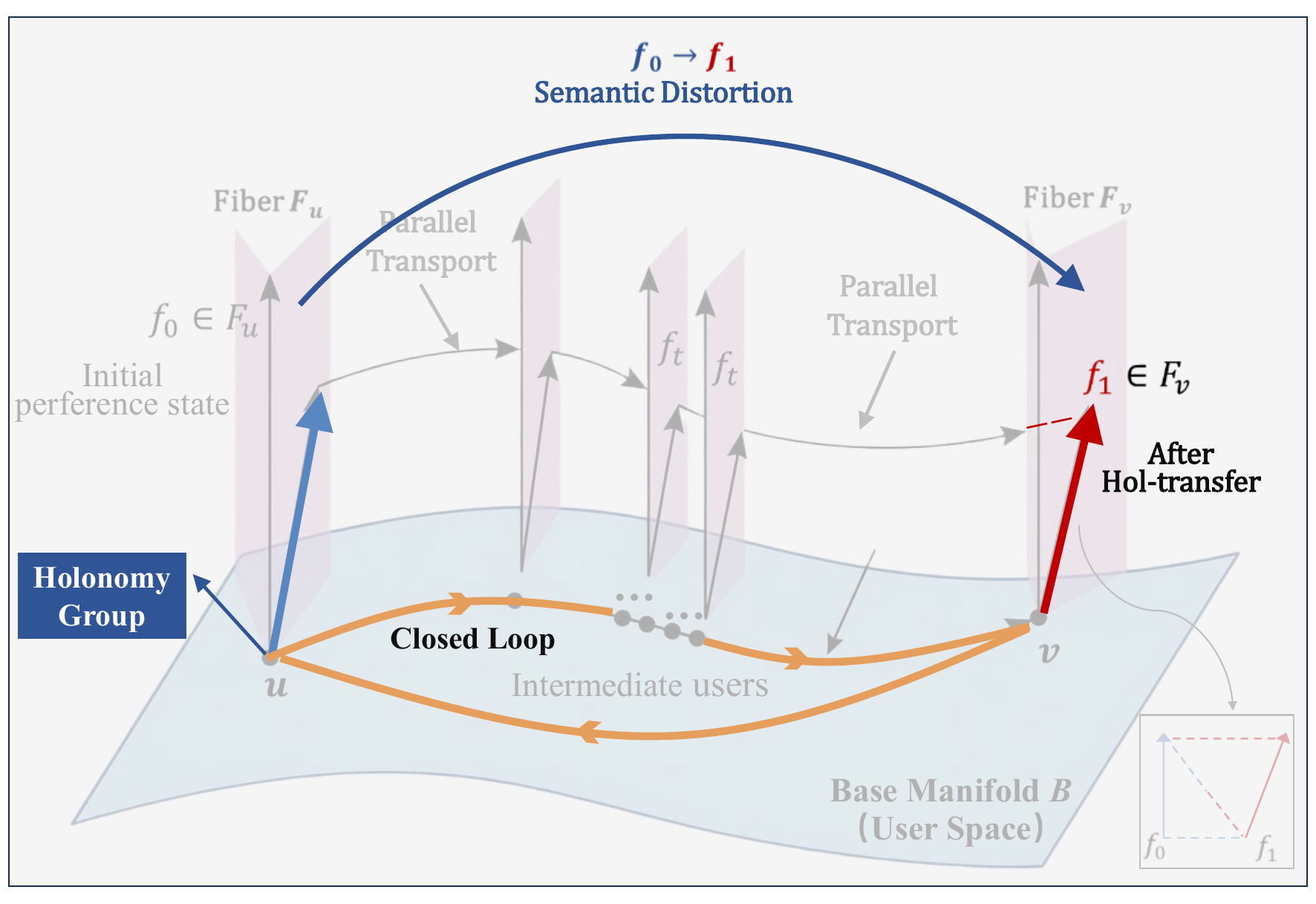}
  \caption{Preference evolution (holonomy transformation)}
  \label{fig:4b}
\end{subfigure}
\caption{Core geometric operators in RecBundle. (a) Parallel transport aligns heterogeneous neighbor features across the base manifold. (b) Closed feedback loops induce holonomy transformations.}
\label{fig:4}
\end{figure*}

Horizontal collaboration in recommender systems aims to leverage similar users to supplement target user information, e.g., neighborhood aggregation in GNNs. From the fiber bundle perspective, since the preference features of users $v$ and $u$, denoted as $f_v$ and $f_u$ respectively, reside in mutually independent fiber spaces $F_v$ and $F_u$, performing direct addition or inner product operations on them in Euclidean space is generally geometrically ill-posed. This represents a primary cause of the coupling of heterogeneous signals.
To achieve rational information collaboration, recommendation algorithms essentially learn a discrete geometric connection $\omega$ implicitly. The aggregation process is equivalent to utilizing a parallel transport operator $P_{v \rightarrow u}$ to transport the preference state $F_v$ of neighbor $v$ along the base manifold and project it into the tangent space of the target user $u$, as illustrated in Figure~\ref{fig:4a}.
\begin{equation}
    P_{v \rightarrow u}(f_v) \approx \alpha_{uv}\mathbf{W}f_v.
\end{equation}

In this discretized representation, the weight matrix $\mathbf{W}$ corresponds to the directional component of the connection, responsible for transforming the feature bases, while the attention coefficient $\alpha_{uv}$ corresponds to the intensity component of the connection.

 When local interactions are sparse or user heterogeneity is strong, the local connection curvature $\Omega$
of the base manifold is typically high. In such cases, enforcing cross-region parallel transport to fit global objectives easily generates significant local spatial distortions. This perspective provides a geometric explanation for why GNNs are prone to over-smoothing \cite{DBLP:conf/aaai/ChenLLLZS20} on tail users and offers a reference for designing robust aggregation mechanisms.

\subsection{Holonomy Transformations}

Besides horizontal collaboration, the other core mechanism of recommender systems is the dynamic closed loop involving interaction-recommendation-feedback. In RecBundle, we view a user's sequential interaction history as a path $\gamma$ on the base manifold $B$. As a user continuously receives recommendations and provides feedback, their initial preference state $f_{init}$ evolves within the fiber space $F_u$. After experiencing one or more feedback loops $\gamma$, influenced by the cumulative effect of algorithmic connections, the state vector undergoes a corresponding linear transformation:$f_{end}=Hol(\gamma) \cdot f_{init}$.
This transformation matrix induced by the closed-loop path is known as the Holonomy, as shown in Figure~\ref{fig:4b}.

From geometric insight, in an ideal setting, $Hol(\gamma)$ should approximate a volume-preserving orthogonal transformation, supporting natural interest migration without dimensional loss. However, under existing algorithms, the holonomy operator manifests as a contraction mapping: spectral decomposition reveals that non-dominant interest components are progressively suppressed during closed-loop evolution. This volume contraction provides a fundamental mathematical characterization of information dynamics in recommender systems.

% \subsection{Discussion}

% RecBundle also serves as a universal geometric perspective for analyzing existing recommendation algorithms. As summarized in Table~\ref{tab:2}, mainstream recommendation paradigms can be viewed as special cases of RecBundle under specific geometric conditions.

\begin{table*}
  \caption{Geometric Interpretations of Mainstream Recommendation Paradigms}
  \small
  \label{tab:2}
   \begin{tabular}{@{}>{\raggedright}m{3.2cm}>{\raggedright\arraybackslash}m{3.2cm}>{\raggedright\arraybackslash}m{2.8cm}m{7.0cm}@{}}
    \toprule
  Algorithm Paradigm & Base Manifold Assumption & Curvature Features & Interpretations \& Limitations from a Geometric Perspective \\
    \midrule
   Matrix Factorization & Euclidean flat space & Zero curvature & Static space; ignores path dependence and state dynamics.\\ 
    Graph Neural Networks & Discrete graph manifold & Implicit linear  & Linearized transport; over-smooths in high-curvature regions.\\
    Large Language Models & Discrete token sequence & Attention weights & Unconstrained paths; prone to semantic hallucinations.\\
    RecBundle (Ours) & Stratified manifold & Explicit holonomy  & Decouples topology \& semantics; quantifies evolutionary bias.\\
    
    \bottomrule
  \end{tabular}
\end{table*}

% In conclusion, introducing fiber bundle theory enables the analysis of the recommendation process through the curvature variations of the base manifold and the spectral characteristics of fiber holonomies. This paradigm shift establishes a solid theoretical foundation for designing future recommender systems with structural robustness and long-term evolutionary capabilities.

\section{Evolution Blueprint}

Building on the RecBundle theoretical framework established earlier, this section demonstrates how this geometric paradigm can address fundamental challenges in recommender systems. We outline its application potential from three perspectives: explainable quantification of information evolution, adaptive recommendation mechanisms, and a novel inference architecture that integrates large language models.

\subsection{An Explainable Framework for Information Evolution}

Most existing research on recommender system explainability relies on post-hoc attribution or static weight analysis. These approaches fail to capture the underlying mechanisms of continuous information evolution \cite{bai2025ensuring, liu2025alleviating, DBLP:conf/aaai/KhooCQ020, DBLP:conf/aaai/BianXXZHRH20}. RecBundle introduces a process-oriented explainability paradigm that unifies three progressively severe phenomena of information degradation: recommendation bias \cite{JSYJ202210001, he2025relieving, krause2024mitigating}, filter bubbles \cite{sunstein2006infotopia, pariser2011filter, Sunstein2003republic}, and rumor propagation \cite{wei2022modeling, jankowski2020evaluation}. The framework integrates these phenomena into a single mathematical formulation based on geometric invariants, specifically curvature and holonomy. It maps these phenomena onto directional shifts and volumetric contraction in the representation manifold, enabling white-box analysis of system dynamics.

Consider information cocoons as an example. Geometrically, information cocoons are characterized as volumetric shrinkage in the feature space. This quantification begins at the local interaction level. When the model aggregates heterogeneous features, it introduces geometric distortion captured by the connection curvature $\Omega$. For discrete graphs, we approximate $\Omega$ using local feature alignment residuals:
\begin{equation}
    \hat{\Omega}_u = \frac{1}{\mathcal{N}(u)} \sum_{v \in \mathcal{N}(u)}\alpha_{uv} \cdot \lVert \mathbf{f_u} - \mathbf{W} \mathbf{f}_v \rVert_2.
\end{equation}
This computation relies solely on the model's native latent states, so $\Omega$ is strictly differentiable. A high $\Omega$ value indicates approximation errors when modeling discrete preference transitions. As local interactions accumulate along feedback paths $\gamma$, long-term evolution is governed by the holonomy matrix $\mathbf{H}_{\gamma}$, defined as the ordered product of update Jacobians $\mathbf{J}_t$. We evaluate the spectral radius $\rho =max \vert \lambda_k \vert$ to measure maximum semantic contraction. We then introduce the Geometric Bias Index (GBI) to capture global volume loss:
\begin{equation}
    GBI=1-\frac{1}{d}\sum_{k=1}^{L} \vert \lambda_k(\mathbf{H}_{\gamma}) \vert,
\end{equation}
where $\lambda_k$ represents the k-th eigenvalue of the holonomy matrix, and $d$ is the feature dimension of the fiber space. As GBI approaches 1, it indicates that numerous orthogonal feature dimensions have been compressed during evolution, leading to severe degradation of the semantic space.

\begin{figure}[tp]
\centering
\includegraphics[scale=0.2]{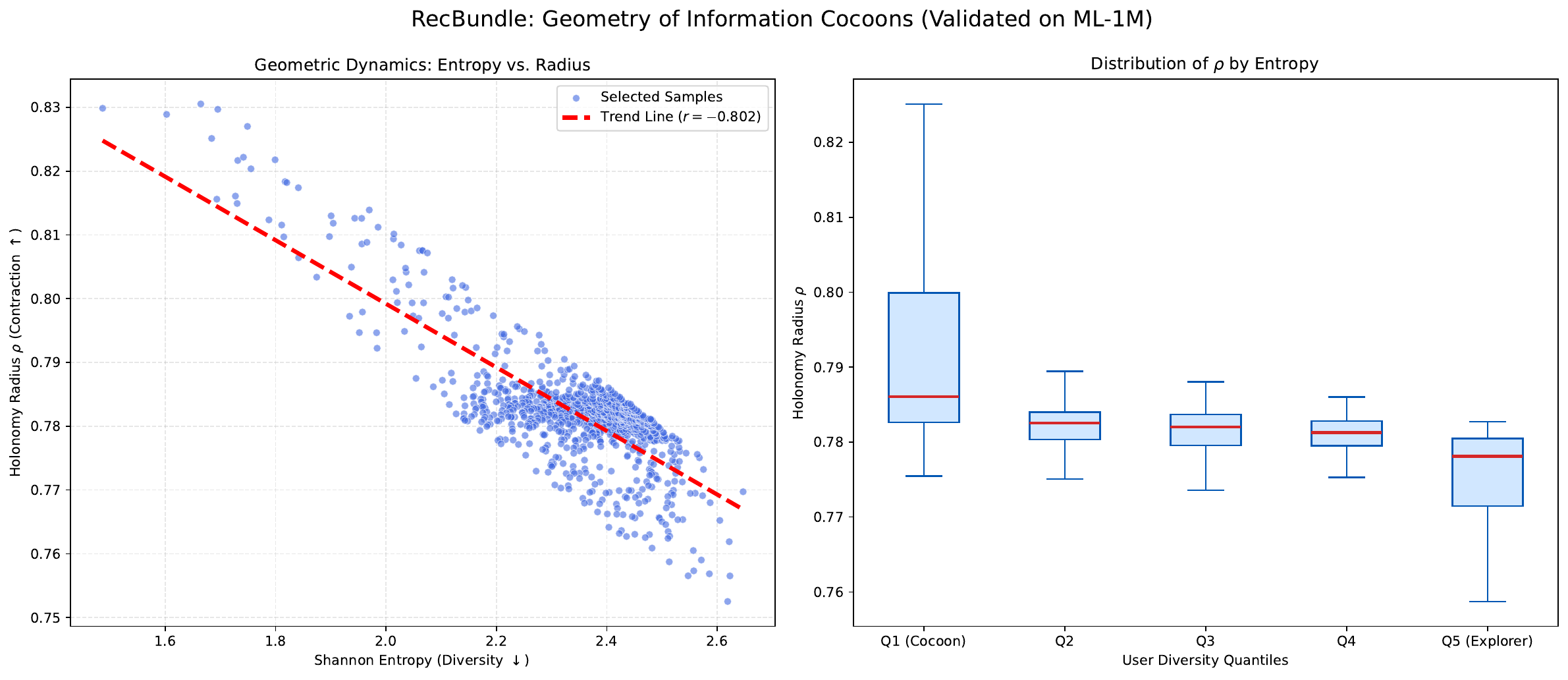}
\caption{The correlation between Shannon Entropy and spectral radius.} \label{Fig_correlation}
\end{figure}

The proposed metrics enable quantitative evaluation in practical recommendation tasks. As shown in Figure~\ref{Fig_correlation}, GBI exhibits a strong negative correlation with Shannon entropy, confirming the consistency between our geometric framework and traditional diversity metrics. Furthermore, Table~\ref{tab:geometric} provides architectural insights at two levels. At the data level, sparsity serves as a primary catalyst for structural contraction , such as Beauty dataset, leading to sharp increases in both $\Omega$ and $\rho$. At the model level, under sparse conditions, architectures with bidirectional dependencies such as BERT4Rec demonstrate significantly higher curvature $\Omega$ compared to unidirectional models such as SASRec. This reveals that complex attention mechanisms are more susceptible to alignment errors when observation samples are limited, making them more prone to information cocoons.

\begin{table}[!htb]
\centering
\caption{Geometric features across real world datasets}
\small
\label{tab:geometric}
% \resizebox{\linewidth}{!}{%
\begin{tabular}{lcccccc}
\toprule
\multirow{2}{*}{Models}  & \multicolumn{3}{c}{ML-1M} & \multicolumn{3}{c}{Beauty} \\
\cmidrule(lr){2-4} \cmidrule(lr){5-7}
 & $\Omega$ & $\rho$ & GBI & $\Omega$ & $\rho$ & GBI \\
\midrule
 SASRec &  0.2514 & 0.1432 & 0.1130 & 0.4534 & 0.3278 & 0.0780 \\
BERT4Rec &  0.3760 & 0.1225 & 0.0870 & 0.9708 & 0.3787 & 0.0911 \\
\bottomrule
\end{tabular}%
\end{table}

This quantification logic extends naturally from information cocoons to rumor propagation. When local geometric deviations accumulate across multiple propagation nodes, they induce substantial directional shifts in feature vectors. This continuous amplification mechanism explains how misinformation exploits network topology to bypass standard semantic constraints.

To address these issues, future optimization can directly regularize specific parameters to maintain stable evolution. First, curvature regularization can optimize attention weights and projection matrices: $\mathcal{L}_{total}=\mathcal{L}_{task}+ \lambda \cdot \frac{1}{\vert U \vert} \sum_{u \in U} \hat{\Omega}_u$, suppressing high-curvature regions at their source. Second, holonomy constraints can regularize sequence encoders by imposing volume preservation conditions ($\vert \det(\mathbf{J}_t) \vert \approx 1$) or spectral penalties on update Jacobians, forcing temporal parameters to preserve the full dimensionality.

\subsection{Metatheory for Adaptive Recommendation}
The Fiber Bundle framework provides a unifying geometric language for adaptive learning systems, revealing that a broad class of meta-learning algorithms \cite{finn2017model,lee2019melu, lu2020meta, luo2020metaselector, dong2020mamo, oreshkin2018tadam, rajeswaran2019meta} can be understood as special cases of this structure. In this view, the base manifold encodes the space of meta-parameters, while each fiber attached to a base point contains task-specific parameters reachable via adaptation. The inner-loop update process is geometrically interpreted as parallel transport along the fiber, with different meta-learning methods corresponding to distinct choices of the connection that governs this transport. For instance, standard MAML with full differentiation implicitly learns a connection capturing second-order curvature \cite{finn2017model}, whereas first-order approximations assume a trivial connection, and methods like iMAML impose a Levi-Civita-like connection through explicit regularization \cite{rajeswaran2019meta}.

Beyond these longitudinal relationships between meta and task parameters, many real-world scenarios exhibit transverse relationships among tasks themselves, such as in federated learning or cross-domain transfer. This motivates extending the fiber bundle to a principal bundle where a structure group acts on fibers, encoding task correlations; methods like HSML that cluster tasks effectively learn such group actions \cite{yao2019hierarchically}, and the resulting holonomy quantifies task interdependence via curvature. Future work will focus on learning bundle geometry end-to-end from data, introducing curvature regularization to enhance knowledge transfer and generalization across tasks, and extending the framework to multimodal scenarios, thereby laying a rigorous theoretical foundation for designing next-generation adaptive algorithms.

% Personalized recommendation must adapt to users' dynamic preferences in non-stationary environments. Traditional approaches rely on local parameter adjustments for different users, forcibly fitting all interaction patterns into a single geometric space. This inherently limits model expressiveness. RecBundle proposes that next-generation adaptive recommendation should evolve from local parameter fine-tuning to dynamic geometric matching.

% User preference evolution involves both exploration and exploitation. During exploration, users exhibit broad semantic interests, requiring low manifold curvature to support long-step feature transitions. During exploitation, users focus on specific domains, where the local space becomes highly structured and nonlinear, requiring stronger connection constraints to preserve fine-grained semantics. The key insight is that algorithms should perceive the local geometric properties of different cognitive phases. Future adaptive frameworks should transcend fixed metric assumptions, dynamically allocating appropriate manifold curvature and connection operators to users at different evolutionary states. This theory provides a unified geometric design principle for complex scenarios such as cross-domain knowledge transfer and adaptive routing.

\subsection{A Novel Inference Architecture with LLMs}

% \cite{xiao2025fastslowthinkinglargevisionlanguage,slow_R1-searcher,slow_du2025virgo11,slow_Slow_Thinking_with_LLMs_1,slow_zhang2025slowthinkingsequentialrecommendation}

As large language models reshape recommender systems, recommendation tasks are evolving from pattern matching toward deep semantic reasoning. State-of-the-art LLMs increasingly explore sophisticated test-time computation and slow thinking mechanisms \cite{du2025virgo, dong2025enhancing, jiang2024enhancing, xiao2025fastslowthinkinglargevisionlanguage, zhang2025slow}. However, without explicit structural constraints, LLMs' processing long interaction sequences often deviates from users' interest boundaries, producing factual errors or logical discontinuities known as semantic hallucinations.

The hierarchical geometry of RecBundle provides a natural framework for integrating LLMs reasoning. Cross-node routing on the base manifold $B$ corresponds to System 1 fast retrieval, using the user-item topological network to bound information extraction. Transitions along a user fiber $F_u$ correspond to System 2 deep chain-of-thought reasoning. Each autoregressive generation step can be formalized as local parallel transport along the fiber. Translating geometric invariants into decoding constraints, such as incorporating curvature as regularization in attention mechanisms or constraining transformation matrices to maintain local smoothness, mechanistically guides the reasoning trajectory. This ensures semantic consistency across multi-step iterations. This paradigm points toward more trustworthy generative recommendations.

\section{Conclusion}
Grounded in the physical and mathematical boundaries of information flow in recommender systems, this work introduced RecBundle, the first geometric paradigm to provide a unified mathematical language for understanding information evolution in complex environments. Building on this geometric foundation, we identify three future directions. First, geometric priors can be integrated as differentiable constraints during training to improve generalization for long-tail users and cold-start items. Second, constructing smooth mappings between heterogeneous manifolds enables knowledge transfer across domains and modalities. Third, structure-aware agents combining local perception with reinforcement learning can dynamically adjust exploration-exploitation trade-offs.

% in response to interest drift.

% Building upon this geometric perspective, we identify three principal directions for future research:

% \textbf{Geometric Prior-Driven Representation Learning}. Transforming spatial geometric characteristics into differentiable regularization constraints and integrating them into the training processes of deep models or large language models endows these systems with topological awareness at the representation learning stage, fundamentally enhancing generalization performance for long-tail users and cold-start items.

% \textbf{Knowledge Transfer Across Heterogeneous Spaces}. Constructing smooth mappings that bridge distinct manifold structures enables high-fidelity knowledge transfer across domains, modalities, and dynamic scenarios, addressing core challenges including semantic gaps, representational heterogeneity, and distribution drift at the structural level.

% \textbf{Structure-Aware Adaptive Decision Agents}. Designing autonomous systems capable of perceiving local structural variations and integrating them with reinforcement learning frameworks allows adaptive adjustment of exploration-exploitation strategies in response to interest drift, popularity fluctuations, and emerging content, thereby enhancing the online robustness of systems operating in dynamic environments. 

We hope RecBundle will inspire a fundamental re-examination of the dynamic mechanisms underlying recommender systems, fostering next-generation paradigms that unite theoretical interpretability with long-term robustness.


% \begin{acks}
% This research was funded by the National Key Research and Development Program of China (Grant No. 2024YFC3307400), under the "Social Governance and Smart Society Technology Support" key special project.
% \end{acks}

\bibliographystyle{ACM-Reference-Format}
\bibliography{sample-base}

\end{document}